\documentclass[aps,prd,onecolumn,showpacs,floatfix,11pt,nofootinbib]{revtex4}
\usepackage{amsfonts}
\usepackage{eurosym}
\usepackage{amssymb,amsmath}
\usepackage{graphicx,float}
\usepackage{indentfirst}
\usepackage{hyperref}
\usepackage{epsfig}
\usepackage{setspace}
\usepackage{graphics}

\usepackage{makeidx}
\usepackage{amsfonts}
\usepackage[ansinew]{inputenc}
\usepackage[usenames,dvipsnames]{pstricks}
\usepackage{subfigure}

\newcommand{\be}{\begin{equation}}
\newcommand{\ee}{\end{equation}}

\newcommand{\ba}{\begin{eqnarray}}
\newcommand{\ea}{\end{eqnarray}}

\begin{document}

\title{\textbf{Refining thick brane models via electroweak data}}
\author{A. E. R. Chumbes$^{1}$}
\email{aruedachumbes@gmail.com}
\author{J. M. Hoff da Silva$^{2}$}
\email{hoff@feg.unesp.br}
\author{M. B. Hott$^{3}$}
\email{hott@feg.unesp.br}
\affiliation{$^{\mathbf{1}}$ Departamento de F\'{\i}sica, UFMA - Universidade Federal do Maranh\~{a}o, \\
Campus Universit\'{a}rio do Bacanga, 65085-580, S\~{a}o Lu\'{\i}s, Maranh\~{a}o, Brazil.\\
$^{\mathbf{2,3}}$
Departamento de F\'{\i}sica e Qu\'{\i}mica, UNESP -
Universidade Estadual Paulista, Av. Dr. Ariberto Pereira da Cunha, 333,
Guaratinguet\'{a}, SP, Brazil}
\date{\today}


\begin{abstract}
After discussing the localization of Abelian and non-Abelian gauge fields
and Higgs fields on a thick brane, we introduce a procedure of dimensional
reduction and its consequences to the rescaled parameters of the boson
sector of the Standard Model. The parameters encodes some power dependence
on the extra dimension, usually narrow, warp factor and hence it also depend
on the position related with the extra dimension inside the thick brane. In
this vein, the observable parameters may be used to refine the braneworld
models via the brane thickness.
\end{abstract}
\maketitle

\section{Introduction}
The investigation of extra dimensions in the warped braneworld context has
attracted a huge attention of the scientific community since the appearance
of the seminal works by L. Randall and R. Sundrum in 1999 \cite{RS1,RS2}. In
both of these models the brane(s) is (are) performed by an infinitely thin
hypersurface(s), and the Standard Model fields are assumed to be trapped on
one of the branes, the so-called visible brane. Such a setup allows a huge
application of the powerful geometrical techniques \cite{JP, ALI}, leading
to a broader understanding of the braneworld paradigm \cite{MAR}.

Soon after the warped braneworld model impact, it was recognized that apart
from the usual expectation of a smooth brane thickness, an infinitely thin
braneworld would give rise to a non negligible quark-lepton operator \cite%
{ARK}, leading to the wrong prediction of, for instance, appreciable rates
for the proton decay. In order to circumvent this problem the consideration
of thick branes has to be taken into account. An heuristic argument in favor
of thick branes goes as follows: let the five-dimensional action for the $%
QQQL$ (Quark-Quark-Quark-Lepton) operator be
\begin{equation}
S\sim \int d^{5}x\sqrt{g}(QQQL).
\end{equation}
By supposing the fields live on the brane, it was found [6] that $Q\sim
e^{-\rho ^{2}r^{2}}q(x^{\mu })$ and $L\sim e^{-\rho ^{2}(r-\Delta
)^{2}}l(x^{\mu })$ where $r$ is the extra dimension, $\rho $ is a parameter
of order of the four-dimensional fermion mass scale and $\Delta $ accounts
for the localization of the quarks and leptons wave functions at different
places within the brane. Therefore, the $\Delta $ parameter brings
information about the brane thickness in this simplified argument. By taking
into account a simple gaussian warp factor $e^{-2\tau ^{2}r^{2}}$ (where $%
\tau $ is a parameter with units of $[length]^{-1}$ usually related to the
inverse of the AdS radius, just for the purposes of this discussion) in
order to guarantee a thick brane scenario, one arrives at
\begin{eqnarray}
&&\hspace{-1.0cm}S\sim \sqrt{\frac{\pi }{2(2\tau ^{2}+\rho ^{2})}}\exp {\Bigg[-\Delta
^{2}\rho ^{2}\Bigg(1-\frac{1}{2(1+\tau ^{2}/\rho ^{2})}\Bigg)\Bigg]}\int d^{4}x\;(qqql).
\end{eqnarray}
In this vein, the existence of some thickness is responsible for attenuate
the effects of the quantum corresponding operator. The resulting (small)
effective coupling constant fixes the problem. More than that, the idea of
universal extra dimensions, in which all standard model fields are present,
has spread over the braneworld models
\cite{cvetic1,wise,townsend,dewolfe,csaki0,
csaki,dewolfe2,gremm,giovannini,campos,ESCA1,ESCA2,koley,melfo,REV,china0,
dutrahottamaro,almeida1,china,aug1,aug2,cazuza,ESCA3}. The combination of
these two characteristics --- universal extra dimensions and thick branes
--- appears, then, as a tool for constructing new models. The brane, within
this context, is understood as a domain wall generated by the maximum slope
region of one or more classic background scalar fields, and this set of
scalar fields also are coupled to gravity
\cite{townsend,dewolfe,csaki0,csaki,dewolfe2,gremm,giovannini}.
The localization of the standard model fields is thus an important issue to
be taken into account in these scenarios, since they must be placed on the
brane in order to accomplish the physical phenomena. This program presents
some recursive patterns. First of all, the scalar field can always be
localized by means of the gravitational weight, in a manner of speaking,
coming from the metric determinant in the effective action \cite{gab}.
However, this is no longer true for the Abelian gauge field. More precisely,
the metric determinant will be of no help in the localization process for
any two-form field term appearing in the Lagrangian. Hence, it is also a
problem for the non-Abelian field strength.

Some important effort has been done for the localization of gauge fields on
the brane \cite{ESCA3,GB1,GB2,almeida,chineses}. Among them, we have
proposed a recursive method based upon analogies to the effective coupling
of neutral scalar field to electromagnetic field and to the Friedberg-Lee
model for hadrons \cite{Noi}. The approach in \cite{Noi} is simply to write
down the action with the kinetic term endowed with a smearing out function,
say $G(\phi )$, where in $\phi $ the classic background scalar field which
give rise to the brane. In practice, every normalizable and symmetric in the
extra-dimension $G$ function may be used for the localization purposes, but
the investigation of a physical model supporting this idea is insightful,
even in the generalization (and its consequences) we shall investigate in
this manuscript. In fact, a possible explanation of the observed $\pi
^{0}\rightarrow \gamma +\gamma $ decay is given in terms of effectively
coupling between a pseudoscalar and a gauge field \cite{SCW}. In \cite{SCW}
it was also developed an effective model that describes the decay of
stationary neutral scalar meson into two parallel polarized photons, where
the usual Maxwell term is coupled with the scalar field as $\phi F^{\mu \nu
}F_{\mu \nu }$. Moreover, in trying to explain low energies QCD
nonperturbative effects, it was proposed by Friedberg and Lee another
effective coupling between a (phenomenologically motivation) scalar field
functional and the gluon kinetic term $f(\phi )\mathcal{F}_{a}^{\mu \nu }%
\mathcal{F}_{\mu \nu }^{a}$ \cite{FL}. After all, the boundary conditions
used in \cite{FL} are also appropriate to the gauge field localization (for
details, see \cite{Noi}).

Given the very nature of the Friedberg-Lee model, its application to the
non-Abelian field localization seems to be somewhat direct. In fact we
consider the localization of zero modes of the non-Abelian gauge field very
much like in the Abelian gauge field, that is, we take as the starting point
the zero mode as a constant and show that this is a possible solution even
when the self-interaction is taken into account. Most importantly, its
possible consequences appear to be relevant to refine braneworld models.
For instance, in Ref. \cite{ult} the study of effective grand unification
scale magnetic monopoles in braneworld slice has enabled the identification
of a subregion of the parameter space in which the model is well defined.
As a matter of fact, when investigating effective models at a fixed extra
dimensional point the resulting effective coupling constants and fields are
dressed by some factors depending on the extra dimensional coordinate. Among
these factors we highlight the warp factor and the smearing out functions.
It turns out that some observables are also dressed by typical combinations
of these factors. Therefore, it is possible to relate some precise
measurements (and their respective errors) with an allowed range for the
brane thickness in a given context.

As we shall see, in the aforementioned framework it is possible to use
electroweak data to refine braneworld models, using the constraints over the
brane thickness in, essentially, two measurements, namely: the Higgs boson
mass and the Weinberg angle. From the Higgs boson effective mass it is
relatively simple to get information (boundaries) on the brane model. For a
specific example, it is shown how to constrain the space of parameters of
the model, but the current data are less stringent than the Weinberg angle
measurements. On the other hand, the boundaries coming from the Weinberg
angle data are more indirect and depend on additional assumptions concerning
the gauge fields localization.

The paper is organized as follows: in section II we discuss the localization
of the electroweak bosonic sector on the brane and perform the usual
procedure of dimensional reduction in order to reproduce the effective
action of the electroweak bosonic sector in $3+1$ dimensions. In section III
we discuss another possible dimensional reduction where the fields are
considered in a four-dimensional slice of the brane, in such a way that the
parameters of the electroweak bosonic sector bear a dependence on the point
of the extra dimension where the slice is located at. Then, we proceed to an
analysis on the dependence of the parameters on the extra dimension and
compare the results with the experimental ones namely, the measured Higgs
mass and Weinberg angle, such that one can refine models of thick branes by
having in mind experimental results. The third section is left to further
comments on our approach and to the conclusions.

\section{Localizing fields on the brane}

This section is devoted to recall some important steps concerning the gauge
and scalar fields localization issue. The final results here will be used to
analyze the effective model and its consequences to the braneworld scenario.

\subsection{\textbf{The abelian case}}

Let us start reviewing the main aspects of the Abelian gauge field
localization \cite{Noi}. Throughout this paper we assume the gravitational
background as
\begin{eqnarray}
ds^{2}=g_{MN}dx^{M}dx^{N}=e^{2A(r)}\eta _{\mu \nu }dx^{\mu }dx^{\nu
}-dr^{2}, \hspace{1.0cm} M,N=0,...,4,  \nonumber  \label{1}
\end{eqnarray}%
being $\eta _{\mu \nu }$ the four dimensional Minkowski metric and $%
e^{2A(r)} $ is the warp factor, which is supposed to depend only on the
extra dimension $r$. The Greek indices run from 0 to 3. Let $\bar{\phi} $ be
the classic background scalar field configuration giving rise to the brane
(therefore a solution of the coupled gravitational equations).

The localization of the gauge field, $V_{\mu }$, by means of the smearing
out function starts with the action
\begin{equation}
S=-\frac{1}{4}\int d^{5}x\sqrt{g}G(\overline{\phi }(r))\mathcal{F}^{MN}%
\mathcal{F}_{MN},  \label{2}
\end{equation}%
where $\mathcal{F}_{MN}=\partial _{\lbrack M}\mathcal{V}_{N]}$. In the
effective model supporting the smearing out function idea, the contribution
of $G(\bar{\phi}(r))$ to the background is neglected. As usual, considering
the $\partial _{\alpha }\mathcal{V}^{\alpha }=0$ and $\mathcal{V}_{4}=0$
gauge, the decomposition $\mathcal{V}_{\mu }(x,r)=\sum\limits_{n}V_{\mu
n}(x)\alpha _{n}(r)$ leads to the following equation of motion for $\alpha
_{n}(r)$
\begin{equation}
m_{n}^{2}\alpha _{n}(r)+e^{2A(r)}\Bigg\{\alpha _{n}^{\prime \prime }(r)+%
\Bigg(\frac{G^{\prime }(\overline{\phi }(r))}{G(\overline{\phi }(r))}%
+2A^{\prime }(r)\Bigg)\alpha _{n}^{\prime }(r)\Bigg\}=0,
\end{equation}%
which after the transformation $\alpha _{n}(r)=e^{-\gamma (r)}g_{n}(r)$,
followed by the identification $2\gamma ^{\prime }=2A^{\prime }+G^{\prime
}/G $ reduces to $-g_{0}^{\prime \prime }(r)+\{\gamma ^{\prime \prime
}+\gamma ^{\prime }{}^{2}\}g_{0}(r)=0$, for the massless zero mode. Now it
is easy to see that $g_{0}\sim e^{\gamma }$, and, as a consequence, $\alpha
_{0}$ is a constant. By considering only the localization of the zero mode
one has
\begin{equation}
S=-\frac{1}{4}\underbrace{\int_{-\infty }^{+\infty }\alpha
_{0}^{2}G(\overline{\phi }(r))dr}_{=1} \int d^{4}xF^{\mu \nu }F_{\mu \nu },
\label{3}
\end{equation}
being $F_{\mu \nu }=\partial _{\lbrack \mu }V_{\nu ]}$, with $V_{\nu }$
standing for $V_{\nu 0}$ for simplicity.

In reference \cite{Noi} it was discussed how to implement a physically
motivated smearing out function. For the purposes of this work we shall just
call attention to the general behavior of $G(\overline{\phi }(r))$.
Therefore, by the very necessity of a convergent function, it is necessary
that $\limsup {G(\overline{\phi }(r))}=const.$ at the brane core ($\bar{\phi}%
(0)$), and $G(\overline{\phi }(r))\rightarrow 0$ as $\bar{\phi}(\pm \infty )$.
Finally, as argued in reference \cite{Noi}, the universal coupling of the
gauge field to matter is respected by the smearing out function procedure
and, hence, the zero modes of all independent fermion fields couple with
equal strength to the zero mode of the gauge field.

\subsection{\textbf{The non-Abelian case}}

In order to proceed with our analysis it is necessary to look at the
non-Abelian gauge field. In what follows we present the main steps to this
case. We shall start saying that the smearing out function procedure can be
used in this case as well and leads to the following action
\begin{equation}
S=-\frac{1}{4}\int d^{5}x\sqrt{g}\tilde{G}(\overline{\phi }(r))\mathcal{F}%
^{MN\,a}\mathcal{F}_{MN\,a},  \label{3b}
\end{equation}%
where $\mathcal{F}^{MN\,a}=\partial ^{\lbrack M}\mathcal{W}%
^{N]a}+g_{5}\varepsilon ^{abc}\mathcal{W}^{Mb}\mathcal{W}^{Nc}$. A crucial
aspect of the adopted point of view: as it can be read from equations (\ref%
{2}) and (\ref{3b}) is that we are assuming different smearing out functions
for the Abelian and non-Abelian case. As we shall see in the next section
this difference is completely irrelevant to explore the Higgs boson mass
parameter, but it is fundamental in constraining the brane model via the
restriction coming from the Weinberg angle measurement. There is no
argument, up to our knowledge, in favor of one or other situation, i.e.,
although the recursive pattern in determining the smearing out function,
shown in reference \cite{Noi}, can be used, there is still room for other
possibilities within this scope. From now on, we shall keep our presentation
dealing with different smearing out functions.

Thus, going further, a similar procedure to the one used in the last
subsection may be applied here. We resort to the field expansion $\mathcal{W}%
_{\mu }^{a}=\sum\limits_{n}W_{\mu n}^{a}(x)\beta _{n}(r)$ and to the gauge
conditions $\partial ^{\mu }W_{\mu n}(x)=0$, $\mathcal{W}_{4}^{a}=0$.
Furthermore, we assume that the zero mode $\beta _{0}$ is constant, as for
the free theory. Parenthetically, we notice that in trying to localize the
full theory (without the constraint $\beta _{0}$ constant), one would face
the intricate problem caused by the term $\sum\limits_{n}\sum\limits_{m}\int
dre^{2A(r)}\tilde{G}(\bar{\phi}(r))(\partial _{r}\beta _{0})^{2}\int
d^{4}xW_{a n}^{\mu }W_{\mu m}^{a}$, from which we can obviously get rid of
by assuming $\beta _{0}$ constant. Then, one has
\begin{eqnarray}
\hspace{-0.1cm}&&S =-\frac{1}{4}\int_{-\infty }^{\infty }dr\beta _{0}^{2}\tilde{G}(%
\overline{\phi }(r))\int \left. \!\!\!\!\!\!\!\!\!\!\hspace{0.5cm}d^{4}x%
\Big\{\partial ^{\lbrack \mu }W^{\nu ]a}\partial _{\lbrack \mu }W_{\nu
]}^{a}+ 2\beta _{0}g_{5}\varepsilon ^{abc}\partial ^{\lbrack \mu }W^{\nu
]a}W_{\mu }^{b}W_{\nu }^{c}+ \right. \nonumber \\
&&\hspace{+0.8cm}+ \left.\beta _{0}^{2}g_{5}^{2}\varepsilon ^{abc}\varepsilon ^{ade}W^{\mu
b}W^{\nu c}W_{\mu }^{d}W_{\nu }^{e}\Big\},\right.  \label{4}
\end{eqnarray}%
which amounts to
\begin{equation}
S=-\frac{1}{4}\underbrace{\int_{-\infty }^{+\infty }\beta
_{0}^{2}\tilde{G}(\overline{\phi }(r))dr}_{=1}\int d^{4}xF^{\mu \nu a}F_{\mu \nu
a},  \label{5}
\end{equation}%
where $F_{\mu \nu \,}^{a}=\partial _{\lbrack \mu }W_{\nu ]\,}^{a}+\beta
_{0}g_{5}\varepsilon ^{abc}W_{\mu \,}^{b}W_{\nu \,}^{c}$, $g_{5}$ is the
non-Abelian coupling constant in five dimensions and $W_{\nu }^{a}$ means $%
W_{\nu ~0}^{a}$. Again, a narrow bell shaped $\tilde{G}(\overline{\phi }(r))$
function would lead to the localization of the non-Abelian gauge field, and
in this dimensional reduction the coupling constant in $3+1$ dimensions $g$
is related to $g_{5}$ by $g=g_{5}\beta _{0}$.

\subsection{\textbf{The Higgs field case}}

In this section we analyze the localization of a complex scalar field on the
brane even when the Higgs potential is taken into account. The action for
the Higgs field coupled with the gravity is
\begin{equation}
S=\int d^{5}x\sqrt{g}\,\left( g^{MN}(D_{N}\Phi )^{\dag }(D_{M}\Phi )-\frac{%
\lambda _{5}}{4}(\left\vert \Phi \right\vert ^{2}-v_{5}^{2})^{2}\right) .
\label{moe}
\end{equation}%
The covariant derivative is $D_{M}=(\partial _{M}+\frac{i}{2}g_{5}\tau ^{a}%
\mathcal{W}_{M\,a}+\frac{i}{2}q_{5}\mathcal{V}_{M})$, where $g_{5}$, $q_{5}$
are respectively the non-Abelian and Abelian coupling constants in five
dimensions. We consider the gauge conditions $\mathcal{W}^{4}=0$, $\mathcal{V%
}^{4}=0$ as in the previous subsections.

As mentioned in the introduction, a scalar field has its zero mode localized
on a brane by means of just the gravitational weight and it is constant as
in the free field case. By using the following expansion $\Phi
(x,r)=\sum\limits_{n}\zeta _{n}(r)\varphi _{n}(x)$ and by taking into
account the interaction among only the zero modes of the Higgs and gauge
fields one finds that (\ref{moe}) can be rewritten as
\begin{eqnarray}
S =\int_{-\infty }^{\infty }dr\zeta _{0}^{2}e^{2A(r)}\int d^{4}x\left(
D^{\mu }\varphi \right) ^{\dag }\left( D_{\mu }\varphi \right)-\frac{1}{4}\int_{-\infty }^{\infty }dr\lambda _{5}\zeta
_{0}^{4}e^{4A(r)}\int d^{4}x\left( \left\vert \varphi \right\vert ^{2}-\frac{%
v_{5}^{2}}{\zeta _{0}^{2}}\right) ^{2},  
\label{6}
\end{eqnarray}%
where $D_{\mu }=\left( \partial _{\mu }+\frac{i}{2}g_{5}\beta _{0}\tau
_{a}W_{\mu \,a}+\frac{i}{2}q_{5}\alpha _{0}V_{\mu }\right) $ and $\varphi
(x) $ stands for $\varphi _{0}(x)$.

Further, we impose that $\int_{-\infty
}^{\infty }dr\zeta _{0}^{2}e^{2A(r)}=1$ and that $\int_{-\infty }^{\infty
}dr\lambda _{5}\zeta _{0}^{4}e^{4A(r)}=\lambda $ and $v^{2}=\left(
v_{5}/\zeta _{0}\right) ^{2}$ turn out to be the Higgs field
self-interaction coupling constant and the parameter of symmetry breaking of
the Higgs potential in $3+1$ dimensions, respectively, such that the
dimensionally reduced effective action for the Higgs fields stands as%
\begin{equation}
S=\int d^{4}x\left( \left\vert D^{\mu }\varphi \right\vert ^{2}-\frac{1}{4}%
\lambda \left( \left\vert \varphi \right\vert ^{2}-v^{2}\right) ^{2}\right) .
\label{7}
\end{equation}%
These redefinitions are possible because we are considering $\zeta _{0}=~$%
\textrm{constant}, as it happens to be in the free scalar field case.
Moreover, from the fact that the zero modes associated to the gauge fields
are also constant we can redefine the coupling constants of the Higgs to the
gauge fields as well, namely $g=g_{5}\beta _{0}$ is the coupling constant of
the Higgs field to the non-Abelian gauge field, whereas $q=q_{5}\alpha _{0}$
is the coupling constant of the Higgs field to the Abelian gauge field in $%
3+1$ dimensions. We notice that the Higgs field couples to the non-Abelian
gauge field with the same coupling constant the non-Abelian gauge fields
couple to themselves; this is a manifestation of the universality of the
charge in this dimensional reduction.

With this procedure of localizing gauge fields and the Higgs field on a
thick brane and also by introducing a consistent dimensional reduction we
are able to reproduce the well-known action for the electroweak bosonic
sector on the brane, namely
\begin{eqnarray}
&&\hspace{-0.8cm}S_{eff}=\int d^{4}x\left\{ -\frac{1}{4}F^{\mu \nu \,a}F_{\mu \nu
\,a}-\frac{1}{4}F^{\mu \nu }F_{\mu \nu }+\left\vert D^{\mu }\varphi
\right\vert ^{2}-\frac{\lambda }{4}(\left\vert \varphi \right\vert
^{2}-v^{2})^{2}\right\} ,  \label{8}
\end{eqnarray}%
where all the information on the extra dimension is encoded in the coupling
constants by their dependencies on the zero modes, which are not observable
quantities.

In the next section we discuss another dimensional reduction
whose consequences on the measurable parameters of the electroweak theory
impose constraints on the brane and on the smearing out functions.

\section{The electroweak bosonic sector on a brane slice}

Here we investigate the consequences of the localization and dimensional
reduction procedure when dealing with the electroweak bosonic sector. In
order to analyze the eventual influence of the brane thickness on some
measurable parameters of the Standard Model we carry out an alternative
procedure for the dimensional reduction of the full action
\begin{eqnarray}
\hspace{-0.5cm}&&S=-\frac{1}{4}\int d^{4}x\int_{-\infty }^{+\infty }dr~\left. \Big\{\alpha
_{0}^{2}G(\overline{\phi }(r))F^{\mu \nu }F_{\mu \nu }+  \beta _{0}^{2}\tilde{G}(\overline{\phi }(r))F^{\mu \nu a}F_{\mu \nu
a} + \right.  \nonumber \\
\hspace{-0.7cm}&& -4\zeta _{0}^{2}e^{2A(r)}\left( D^{\mu }\varphi \right) ^{\dag
}\left( D_{\mu }\varphi \right) +\lambda _{5}\zeta _{0}^{4}e^{4A(r)}\left( \left\vert \varphi \right\vert
^{2}-\frac{v_{5}^{2}}{\zeta _{0}^{2}}\right) ^{2}\Big\}.  \label{8a}
\end{eqnarray}
Instead of integrating the extra dimension we consider all the fields placed
at a four dimensional slice of the thick brane, that is supposed to be
localized at $r=\overline{r}$ and with width $L$.  Notice that in dealing with the bosonic sector in a given slice we shall not struggle with $r$-dependent probabilities as it would be the case for fermionic fields. Instead the bosonic fields and coupling constants are rescaled in a consistent way as follows
\begin{eqnarray}
\overline{V}_{\nu } &=&V_{\nu }\,G(\bar{\phi}(\overline{r}))^{1/2}\,,~%
\overline{W}_{\nu }^{a}=W_{\nu }^{a}\,\tilde{G}(\bar{\phi}(\overline{r}%
))^{1/2},  \notag \\
~\bar{\varphi} &=&e^{A(\bar{r})}\varphi ,\hspace{1.6cm}\bar{q}=q\tilde{G}(%
\bar{\phi}(\overline{r}))^{-1/2},  \nonumber \\
~\bar{g} &=&g\tilde{G}(\bar{\phi}(\overline{r}))^{-1/2},\hspace{0.5cm}%
\overline{\lambda }=\lambda _{5}\zeta _{0}^{2},  \label{9}
\end{eqnarray}%
and the zero modes are conveniently chosen to be given by $\alpha _{0}=\beta
_{0}=\zeta _{0}=L^{-1/2}$. In this way, the dimensionally reduced effective
action at a given slice of the brane reads%
\begin{eqnarray}
\hspace{-0.7cm}&&\bar{S}_{eff} =\int d^{4}x\left( -\frac{1}{4}\overline{F}^{\mu \nu a}%
\overline{F}_{\mu \nu }^{a}-\frac{1}{4}\bar{f}^{\mu \nu }~\bar{f}_{\mu \nu
}+\left\vert \bar{D}^{\mu }\overline{\varphi }\right\vert ^{2}-\frac{\bar{\lambda}}{4}\left( \left\vert \overline{\varphi }%
\right\vert ^{2}-\bar{v}^{2}\right) ^{2}\right) ,  \label{10}
\end{eqnarray}%
where $\bar{D}_{\mu }=\left( \partial _{\mu }+\frac{i}{2}\bar{g}\tau _{a}%
\bar{W}_{\mu \,a}+\frac{i}{2}\bar{q}\bar{V}_{\mu }\right) $ and $\bar{v}%
^{2}=e^{2A(\bar{r})}v^{2}$.

Some comments are in order at this point. First, one can notice that the
universality of the charge is preserved in this dimensional reduction.
Second, there is no dependence of the parameters on the extra dimension, but
at each different slice they would assume different values, that is, from
the point of view of a brane observer, the parameters are effectively
`running' ones. The relevance of the aforementioned behavior is that, again,
the brane thickness will be constrained. We also remark that convergence
issues are safe by means of the previous section discussion. From now on, we shall give a
prescription of how to use the constraint of the brane thickness associated
to experimental data to refine the braneworld models. The idea is quite
simple and it can be implemented in several levels and/or sophistication
degrees.

\subsection{\textbf{Using the Higgs boson mass measurements}}

The fastest way to find constraints over the brane thickness is from the
rescaled Higgs boson mass. In fact, from the effective Higgs potential
symmetry breaking scale we have any measured mass parameter given by $\bar{m}%
=e^{A(\bar{r})}m$. The recent data by the CMS collaboration have shown a
consistent excess of events above the background proton-proton collisions at
$7$ and $8$ $TeV$ center of mass energy. The data points to a scalar
particle with mass around $125GeV$ \cite{CMS}. More precisely, an adequate
fit of the decay modes $\gamma \gamma $ and $ZZ$ is obtained for a mass
given by $125.3\pm 0.4(stat.)\pm 0.5(syst.)$ $GeV$. Therefore, bearing in
mind a narrow shaped warp factor, present in the majority of models, it is
easy to see that
\begin{equation}
124.4\hspace{4pt}GeV\leq e^{A(\bar{r})}m\leq 126.2\hspace{4pt}GeV.
\label{indo1}
\end{equation}%
Moreover, as the warp factor reaches its maximum value at the brane core ($%
\bar{r}=0$) it is possible to write
\begin{eqnarray}
&&\left. e^{A(0)}m=126.2\hspace{4pt}GeV,\right.  \nonumber \\
&&\left. e^{A(r_{+})}m=124.4\hspace{4pt}GeV,\right.  \label{indo3}
\end{eqnarray}%
where $r_{+}$ stands for the brane `surface'.

It is insightful to look at a specific example from the braneworld
scenarios, in order to see how the conditions (\ref{indo3}) can be used to
refine a given model. Briefly speaking, the so-called Gremm's model is given
by a five dimensional domain wall performed by a scalar field coupled to
gravity \cite{gremm}. By using superpotential technique, it was shown a
formally compatible warp factor given by $e^{-b\ln (2\cosh (2cr))}$, where $%
bc$ provides the AdS curvature of the model. Then, from (\ref{indo3}) and
referring to $\delta \equiv 2r_{+}$ as the brane thickness one arrives at
\begin{equation}
\delta =\frac{1}{c}\mathrm{arccosh}[(1.014)^{1/b}].  \label{indo5}
\end{equation}%
It is possible to go further in the analysis by associating lower and upper
boundaries to the brane thickness, as follows: it is quite conceivable to
require that $\delta \geq l_{(5)}$, the five-dimensional Planck length ($%
2.0\times 10^{-19}m$). On the other hand, current experiments dealing with
possible deviations from the inverse-square Newton's law give $\delta
<44\times 10^{-6}m$ \cite{1sobrer2}. Plugging such constraints into (\ref%
{indo5}) it is possible to find a region in the parameter space which
entails a domain for the AdS curvature of the model. The allowed domain is
shown in Fig. 1.

\begin{figure}[h]
\centering
\includegraphics[width=7.0cm]{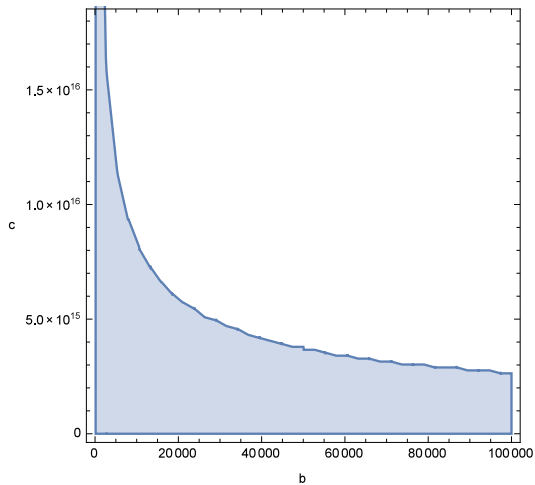}
\caption{(Colour on-line) The parameter space associated to the parameters b
and c, where $0\leq c \leq 2\times10^{16}m^{-2}$ and $0 \leq b \leq 10^{5}$.
}
\label{fig:qqq}
\end{figure}

\subsection{\textbf{Weinberg angle data}}

Here we proceed within the example of the Weinberg angle, a right precise
measured quantity. After the spontaneous symmetry breaking, the perturbative
spectrum can be straightforwardly read from the quadratic terms. Concerning
the gauge fields, the relevant terms come from the covariant derivative $|%
\bar{D}_{\mu }\bar{\varphi}|^{2}$. Assuming the symmetry breaking along the
third isospin component, the right spectrum is reached via the
identification
\begin{equation}
\left(
\begin{array}{c}
\bar{Z}_{\mu } \\
\bar{A}_{\mu }%
\end{array}%
\right) =\left(
\begin{matrix}
\cos \theta _{W} & -\sin \theta _{W} \\
\sin \theta _{W} & \cos \theta _{W}%
\end{matrix}%
\right) \left(
\begin{array}{c}
\overline{W}_{\mu }^{3} \\
\overline{V}_{\mu }%
\end{array}%
\right) ,  \label{11}
\end{equation}%
where $\theta _{W}$ is the Weinberg angle, $Z_{\mu }$ is the massive neutral
boson and $A_{\mu }$ is the electromagnetic field. From the very definition
of the Weinberg angle, it is readily verified that $\tan \theta _{W}=\bar{q}/%
\bar{g}$, which reads as
\begin{equation}
\tan \theta _{W}=\frac{q_{5}}{g_{5}}\frac{\tilde{G}^{1/2}(\overline{\phi }(%
\bar{r}))}{G^{1/2}(\overline{\phi }(\bar{r}))}\equiv \mathcal{G}(r).
\label{12}
\end{equation}%
Now we are in position to use the Weinberg angle value (and its associated
error) to constraint the brane thickness and, then, refining a given model.
For the argument suppose\footnote{%
Actually, the usually measured quantity is $\sin ^{2}\theta _{W}$. Hence, we
may complete the argument by saying that $\sin \theta _{W}=n\pm \Delta n$.
Then,
\begin{equation}
\tan \theta _{W}=\sqrt{\frac{n\pm \Delta n}{1-(n\pm \Delta n)}}.  \nonumber
\end{equation}%
Therefore, the association of $\frac{A\pm \Delta A}{B\pm \Delta B}=\frac{A}{B%
}\pm \frac{A}{B}\Big(\frac{\Delta A}{A}+\frac{\Delta B}{B}\Big)$ and $(A\pm
\Delta A)^{k}=A^{k}\pm kA^{k-1}\Delta A$, leads straightforwardly to $\tan
\theta _{W}=N\pm \Delta N$.} $\tan \theta _{W}=N\pm \Delta N$. Assuming that
the brane core is positioned at $\bar{r}=0$, we shall understand $\mathcal{G}%
(\overline{r})$ with $\overline{r}\in \lbrack r_{-},r_{+}]$ (the extremes of
the brane) and $\mathcal{G}(\overline{r})\in \lbrack N-\Delta N,N+\Delta N]$
as the mapping
\begin{eqnarray}
\mathcal{G}(r) &:&\left. \mathbb{R}\rightarrow \mathbb{R}\right.  \nonumber \\
&&\left. r\mapsto \mathcal{G}(r).\right.  \notag
\end{eqnarray}%
Giving the fact that the brane thickness shall not be macroscopic (by the
reasons previously exposed), it is possible to expand the $\mathcal{G}$ as
\begin{equation}
\mathcal{G}(r_{+})=\mathcal{G}(0)+\frac{d\mathcal{G}(0)}{dr}r_{+}+\frac{1}{2}%
\frac{d^{2}\mathcal{G}(0)}{dr^{2}}r_{+}^{2}+\cdots .
\end{equation}%
To fix ideas, we make the first order term equal to zero, as it is for the
symmetry condition. After a simple algebra we have the following constraint
\begin{eqnarray}
&&8\left( \frac{d^{2}\mathcal{G}(0)}{dr^{2}}\right) ^{-1}(N-\Delta N-\mathcal{G%
}(0))\leq \delta ^{2}\leq 8\left( \frac{d^{2}\mathcal{G}(0)}{dr^{2}}\right)^{-1} (N+\Delta N-\mathcal{G}(0)).
\end{eqnarray}%
Hence the brane thickness is constrained in the following context: for a
given model, whose dependence on the extra dimension is encoded in Eq. (\ref%
{12}), the $\delta ^{2}$ value must respect the numerical restriction coming
from the Weinberg angle measurement. Obviously the analysis is model
dependent, but the point to be stressed is that the above analysis may be
used in order to refine the model itself, constraining its otherwise free
parameters and enabling, thus, the physical viability of the model. Now let
us focusing in another point concerning this reasoning. By implementing the
same boundaries as in the previous analysis, it is fairly trivial to see
that the following inequalities must be fulfilled
\begin{eqnarray}
&&\left. (N-\mathcal{G}(0))\leq 2,4\times 10^{-10}\left( \frac{d^{2}\mathcal{%
G}(0)}{dr^{2}}\right) -\Delta N ,\right.  \notag \\
&&\left. (N-\mathcal{G}(0))\geq 5\times 10^{-39}\left( \frac{d^{2}\mathcal{G}%
(0)}{dr^{2}}\right) +\Delta N ,\right.  \notag
\end{eqnarray}%
therefore
\begin{equation}
\Delta N\left( \frac{d^{2}\mathcal{G}(0)}{dr^{2}}\right) ^{-1}\leq 1,2\times
10^{-10}\left( 1-\frac{10^{-26}}{484}\right)m^{2} .  \label{ufa1}
\end{equation}%
Hence, we see that $\frac{d^{2}\mathcal{G}(0)}{dr^{2}}$ must be positive.
More precisely, disregarding the $10^{-26}$ term, and evaluating the $\Delta
N$ factor (which amounts out to be about $1,2\times 10^{-10}$) we have
\begin{equation}
\frac{d^{2}\mathcal{G}(0)}{dr^{2}}\geq 1,0 m^{-2}.  \label{ufa2}
\end{equation}%
It is interesting to notice, then, that this procedure may also refine the
localization mechanism itself, by means of the smearing out functions. In
fact, by the identification (\ref{12}), we have
\begin{equation}
\left( \frac{\widetilde{G}(0)\mid G^{\prime \prime }(0)\mid -\mid \widetilde{%
G}^{\prime \prime }(0)\mid G(0)}{G^{2}(0)}\right) \geq \frac{g_{5}}{q_{5}}.
\end{equation}%
For instance, we have found under some assumptions in \cite{Noi} that $G(%
\overline{\phi}(r))= \mathrm{sech}^{2k}(2 c r)$ in the case of the Gremm
Model. Similarly, one could choose $\widetilde{G}(\overline{\phi}(r))=
\mathrm{sech}^{2\widetilde{k}}(2 c r)$. This would furnishes, for instance, $%
8 c^{2}(k-\widetilde{k})\geq \frac{g_{5}}{q_{5}}$ for  $k > \widetilde{k}$ and $%
c > 0$, constraint some otherwise free parameters.

\section{Final Remarks}

By investigating the data from effective bosonic electroweak sector of the
standard model in association to current experimental boundaries related to
the brane thickness we were able to indicate useful refinements concerning
the modeling of braneworlds. The idea of considering electroweak data with
respect to their relation to extra dimensions is not new (see, for instance
\cite{APP}), but our approach is essentially the use of the data related to
the rescaled quantities, instead of analyzing radiative corrections.

By using the Higgs boson mass data we find, in particular, for the so-called
Gremm's model, a region in the parameter space $(b,c)$ which serves as
allowed domain to the AdS curvature of the model $bc$. Some similar region
was obtained in \cite{ult} but here we were able to evince, due to the Higgs
mass data, another region of the allowed domain. We have also analyzed the
possible constraints in the brane thickness with respect to the stringent
data coming from Weinberg angle measurements.
The procedure is particularly
interesting whenever the smearing out functions are in place for the gauge
fields localization. Ultimately, this procedure is relevant to constraint
background parameters arising from the thick brane modeling in warped spaces.

\section*{Acknowledgments}
A. E. R. Chumbes thanks to PNPD/CAPES. This work is partially financed by
Conselho Nacional de Desenvolvimento Cient\'{i}fico e Tecnol\'{o}gico/Brasil - CNPq


\end{document}